\documentclass[conference]{IEEEtran}
\usepackage{cite}
\usepackage{url}

\ifCLASSINFOpdf
 \usepackage[pdftex]{graphicx}
\else

\fi

\usepackage{subfigure}

\usepackage{algorithmic}
\usepackage{algorithm}  
\usepackage{color}
\usepackage{xcolor}

\makeatletter
\newcommand\fs@betterruled{%
	\def\@fs@cfont{\bfseries}\let\@fs@capt\floatc@ruled
	\def\@fs@pre{\vspace*{5pt}\hrule height.8pt depth0pt \kern2pt}%
	\def\@fs@post{\kern2pt\hrule\relax}%
	\def\@fs@mid{\kern2pt\hrule\kern2pt}%
	\let\@fs@iftopcapt\iftrue}
\floatstyle{betterruled}
\restylefloat{algorithm}
\makeatother

\makeatletter
\newenvironment{contract}[1][ht]{%
	\renewcommand{\ALG@name}{Contract}
	\begin{algorithm}[#1]%
	}{\end{algorithm}}
\makeatother

\usepackage{amsmath,amsthm,bm,amssymb}
\newtheorem{defi}{Definition}
\newtheorem{prop}{Proposition}

\newtheorem{thm}{Theorem}

\IEEEoverridecommandlockouts
\begin{document}

\title{ Contract-based   Time-of-use Pricing for Energy Storage Investment \vspace{-4mm}}

\author{Dongwei Zhao, Hao Wang,  Jianwei Huang, Xiaojun Lin
\thanks{This work is supported by the Shenzhen Institute of Artificial Intelligence and Robotics for Society, and the Presidential Fund from the Chinese University of Hong Kong, Shenzhen.}
\thanks{Dongwei Zhao is with Department of Information Engineering, The Chinese University of Hong Kong. Hao Wang is with Department of Data Science and Artificial Intelligence, Monash University. Jianwei Huang is with the School of Science and Engineering, The Chinese University of Hong Kong, Shenzhen, and the Shenzhen Institute of Artificial Intelligence and Robotics for Society (AIRS) (corresponding author, e-mail: jianweihuang@cuhk.edu.cn).  Xiaojun Lin is with School of Electrical and Computer Engineering, Purdue University.}}

\maketitle
\begin{abstract} 

Time-of-use (ToU) pricing is widely used by the electricity utility.
A carefully designed ToU pricing can incentivize end-users' energy storage deployment,  which helps shave the system peak load and  reduce the system  social cost.  However, the optimization of ToU pricing is highly non-trivial, and an improperly designed ToU pricing may lead to storage investments that are far from the social optimum. In this paper, we aim at designing the optimal ToU  pricing, jointly considering  the social cost of the utility and  the  storage investment decisions of users.    Since the storage  investment costs are users'  private information, we design  low-complexity contracts to elicit the necessary information and induce the proper behavior of users' storage investment.  The proposed  contracts only  specify three contract items,  which guides  users of  arbitrarily many different storage-cost types to invest in full, partial, or no storage capacity with respect to their peak demands.	Our contracts can  achieve the  social optimum when the utility  knows the aggregate demand   of each  storage-cost type  (but not the individual user's type).  When  the utility only knows the distribution  of each storage-cost type's demand, our contracts can lead to a near-optimal solution. The gap with the social optimum is as small as    1.5\%  based on the simulations using realistic data. We also show that the proposed contracts can  reduce the system social cost  by over 30\%,  compared with no storage investment benchmark.

\end{abstract}

\IEEEpeerreviewmaketitle

\section{Introduction}
\subsection{Background and motivation}
Time-of-use (ToU) pricing is widely used by the electricity utility to shave the system  peak load and reduce the system cost \cite{tous}. With  ToU pricing, the utility divides one day into two or three fixed time periods.  For example, in two-period ToU pricing, the utility sets a higher electricity price  for the peak period (e.g., 4PM to 9PM) and  a lower price for the off-peak period (e.g., 10PM to 3PM)\cite{timeofusey}.  

The price difference between the peak and off-peak periods  provides  incentives for end-users'  energy storage deployment, which can reduce their electricity bill\cite{pimm2018time}.  
Users with  storage can purchase more electricity (by charging the storage) during  the off-peak hours with a lower price. During the peak hours, users can discharge the storage to serve the demand with less electricity consumption from the utility at the higher price.

The increasing deployment of energy storage at end-user side, however,  poses new challenges for the ToU pricing design. On  one hand, a proper ToU pricing can incentivize users to invest in storage and reduce their energy cost, which can further  shave the system peak load and reduce the social cost (compared with no storage investment).  On the other hand,  if the ToU pricing  design does not consider the storage impact and sets the price difference too high,   it may incentivize too much storage investment and  create a new and even higher  system peak load. 
For example, if all the users invest in storage and shift the demand from the peak period to off-peak period, the original peak period will have zero demand while the original off-peak period will become the new peak.   Furthermore, such an excessive storage investment may not be good for the social welfare, as some users may have high storage investment costs.

The above considerations motivate us to study the following problem: \textit{How should we design  the ToU pricing to benefit users who invest in storage and achieve the social optimum  that jointly considers the utility's supply cost and the users' storage investment costs? }  Notice that users can choose different storage products with different technologies, hence users can have heterogeneous storage costs.   
To reach the social optimum, we need to incentivize more users with low storage costs to invest in storage  while discouraging users with high storage costs from investing. However, users' storage costs are  often their private information, which poses challenges for the ToU pricing design. To solve this problem, we will use contract theory to  elicit necessary information of  users and induce the proper storage investment behavior.  

\subsection{Related work}

There have been a substantial amount of works on designing  ToU pricing for the utility.
Chen \textit{et al.} \cite{contract3}  offered optimal contract options of ToU pricing  to  households, which both minimizes the system peak load and maximizes the utility's profit.   K{\"o}k \textit{et al.} \cite{touren} studied optimal ToU pricing with the renewable energy investment. However, these works did not consider the possible impact of  end-users' storage investment. 

There are also works that studied the optimal storage operation and investment under ToU pricing. Nguyen \textit{et al.}\cite{timestorage1} optimized the operation of energy storage to minimize the users' cost under ToU pricing. Carpinelli \textit{et al.} \cite{carpinelli2016probabilistic} proposed a probabilistic method to size the storage  under  ToU pricing.  However, the ToU pricing is exogenously given in these works without considering the storage's impact on the system. So far, there is no literature studying the design of  ToU pricing considering the end-users' storage investment.

In the ToU pricing design, we adopt contract theory to deal with users' private storage costs. Contract theory has been widely used in power systems as the mechanism for  energy or service procurement.  Most of the studies  focused on the optimal contract for maximizing  the payoff of the provider. For example,  Tavafoghi \textit{et al.} \cite{contract1} designed  an optimal contract of  energy procurement for a strategic electricity seller. Haring \textit{et al.} \cite{contract2}  proposed a contract  that incentivizes  users to offer demand response services.  　 Some works also considered the contract  for other objectives.  Chen \textit{et al.}  \cite{contract3} designed  contracts for ToU pricing, which also considered minimizing the system peak load. However, the study in   \cite{contract3} focused on numerical solutions  without theoretical  optimality guarantee.

Different from existing studies, our study aims to analytically design effective yet simple contracts that minimize the social cost. The works  \cite{contract1}  \cite{contract2} designed a different contract item for each type of agents.  However, in our work, we only need to design at most three common contract items for all types of users.  This significantly reduces the complexity of the contract and makes it much more implementation friendly. 

\vspace{-1mm}
\subsection{Main results}
\vspace{-1mm}

Our work focuses on the optimal contract design of the ToU pricing and end-users'  storage investment, which aims to minimize the system social cost.  The main contributions and results of this paper are listed as follows.


\begin{itemize}
	\item \textit{Storage-aware ToU pricing}:  To the best of our knowledge, this is the first work to analytically  study the social-optimal ToU pricing design,  considering  the end-users' storage investment. The increasing storage deployment at the  end-user side requires a new design of ToU pricing to maximize social welfare.
	
	\item \textit{Optimal contract design}:  It is challenging to optimize the  ToU pricing with private user storage costs. We adopt contract theory to solve this problem, and only utilize  the aggregate demand information of each type of storage cost.  
	We analytically design two simple yet effective contracts \textbf{TS-C}  and \textbf{TS-I}, under the complete and incomplete demand information of  types, respectively.  For an arbitrary number  of types with diverse storage costs, we only need  three contract items for each contract.
	
	\item \textit{Contract performance}: 	 We prove that  Contract \textbf{TS-C} can   achieve the  social optimum when the utility knows the aggregate demand
of each type. Based on realistic-data simulations, we show that even when the utility only knows the distribution (but not the exact value) of each type's demand, Contract \textbf{TS-I} can still lead to a near social optimum with a 1.5\% gap with the social optimum.

	\item \textit{Benefits of contract}: Via realistic-data simulation, we show that Contract \textbf{TS-I} can reduce the social cost by over 30\%, compared with a ToU pricing that provides low incentives and leads to no storage investment.  
\end{itemize}

\section{System Model}
  
\begin{figure}[t]
	\centering
	\includegraphics[width=3.1in]{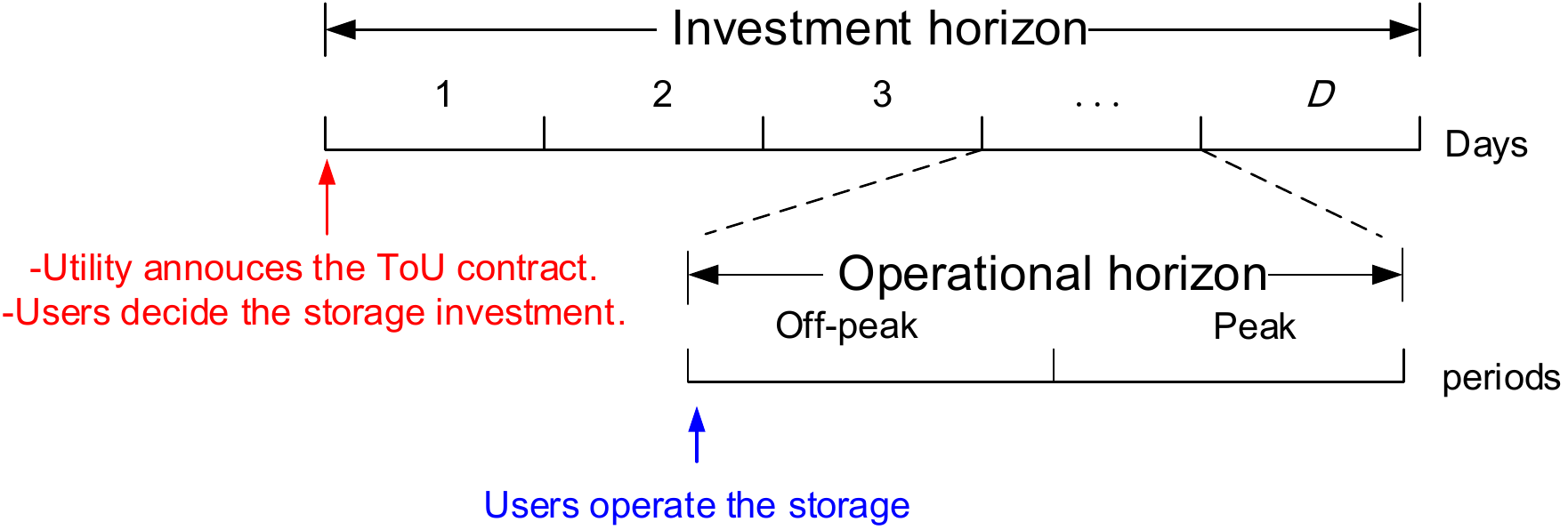}
	\vspace{-2mm}
	\caption{\small Two timescales.}
	\label{fig:time0}
	\vspace{-4mm}
\end{figure}

We consider  one electric utility serving a group  $\mathcal{I}=\{1,2\ldots I\}$ of users (e.g.,  residential users and business users).
Figure \ref{fig:time0} illustrates two timescales of decision making.  At the beginning of an investment horizon 
of $D$ days (e.g., $D$ corresponding to many years), the utility announces the ToU pricing contract to users and users  decide how much to invest in storage. The investment horizon is divided into operational horizons. Each operational horizon corresponds to one day, which is further divided into two periods $\mathcal{T}\hspace{-1mm}=\hspace{-1mm}\{p,o\}$ with peak period $p$ and off-peak period $o$. On each day, each user utilizes  storage to minimize his electricity cost.   Next we will introduce the detailed models of the ToU pricing,  users, and utility.

\vspace{-1mm}
\subsection{ToU pricing }
\vspace{-1mm}

We consider a two-period  ToU pricing\footnote{In practice, both two-period and three-period ToU pricing exist. We focus on the two-period pricing  since it can already leverage the storage's impact, and we leave three-period case as future work.} commonly used by utilities \cite{timeofusey}, which  is announced once and is valid for the entire investment horizon.  For example, 
  the peak period can be set  from
4PM to 9PM and the off-peak period can be from  10PM to 3PM\cite{timeofusey}. We assume that there are  $H^{p}$ hours for the peak period and $H^{o}$ hours for the off-peak period, where $H^{p}+H^{o}=24$.      The electricity prices of the  peak and off-peak periods for user $i$ are  $p_i^p$ and $p_i^o$, respectively. The peak price is no smaller than the off-peak price, i.e.,  $p_i^p\geq p_i^o$.  Notice that we allow the utility to charge different users different prices.

\subsection{Users}
Users face the ToU pricing from the utility.  Based on ToU pricing, users can invest (at the beginning of the investment horizon) and operate the storage (every day) to shift the demand.  For each user $i$, we denote his total demand   in the peak period (of a single day)  as $D_i^p$ and the  demand  in the off-peak period (of a single day)  as  $D_i^o$.  We assume that each user's demand pattern is the same  across different days for analysis  tractability.\footnote{In the more realistic case, a user's demand will vary across days; such a general modeling will significantly complicate our analysis and hence will be left for future work.} Because of this, we do not need to use day index for the demands.  

Next, we first explain the storage cost and electricity bill of users, and then we formulate and solve the  user's cost minimization problem.
\vspace{-0mm}
\subsubsection{Storage cost} At the beginning of the investment horizon, each user $i$ decides the invested storage capacity $c_i$. The unit capacity cost of storage for user $i$  is  $\theta_i$ per day, where we scale the total investment capacity cost of the entire investment horizon  into one day.\footnote{We can evenly divide the total capital cost by the number of days of the investment horizon  considering zero annul interest rate. We also provide a scaling method with  the annul interest rate in the online appendix \cite{project3append}.}  Note that different users can choose different storage products with different technologies, hence users have heterogeneous private storage costs.   User $i$'s total storage cost per day is $\theta_i c_i$. 

\subsubsection{Electricity bill} Next, we calculate the daily electricity bill  for users considering the storage deployment. During each day,    in the off-peak period,  if user $i$ purchase $s_i$ amount of  energy from the utility and charge the storage,\footnote{The payment in ToU pricing is  based on the total consumed energy in peak and off-peak periods, without considering demand variation across hours. Thus, we only let the total charged and discharge energy  be $s_i$. We  assume  users' charge and discharge of storage across hours within peak and off-peak periods can be regulated by the utility \cite{storageprogram2019utility}, so as to smooth the  system load.} the total electricity consumption from the utility will be $D_i^o+s_i$. As a result, in the peak period, the total consumption from the utility will be $D_i^p-s_i\geq 0$.\footnote{We do not consider the negative demand in the current model, i.e., we do not allow users to sell back energy from the storage to the utility \cite{storageprogram2019utility}.}   All the energy  charged into the   storage during the off-peak period will be discharged to serve demand in the peak period. Therefore, at the beginning of the off-peak period of the next day, the initial energy level of storage is zero.
Then, user $i$'s daily electricity bill is  $p_i^p (D_i^p-s_i)+p_i^o (D_i^o+s_i)$.  We consider 100\% charge and discharge efficiency in our model, and we will generalize results  to non-100\% efficiency  in the future journal version.

\subsubsection{User's cost minimization problem} Each user wants to minimize his total cost, which includes the electricity bill  and the cost of storage investment (scaled into one day). If users  are only charged based on the ToU pricing specified in Section II.A,  then  each user $i$'s cost minimization problem is formulated in Problem \textbf{UCM}. Each user $i$ decides the invested storage  capacity $c_i$ over the investment horizon  and the charged energy $s_i$ in the off-peak period each day  (where $s_i$ does not change over different days).

\noindent \textbf{Problem UCM: User $i$'s Cost Minimization }
\begin{align}
\min~ &p_i^p (D_i^p-s_i)+p_i^o (D_i^o+s_i)+\theta_i c_i\notag \\
\text{s.t.~} &0\leq s_i\leq c_i,\label{eq:userc1}\\
&s_i\leq D_i^p,\label{eq:userc2}\\\
\text{var}:~ &c_i, ~s_i. \notag
\end{align} \par \vspace{-2mm}
\noindent  Constraint \eqref{eq:userc1}  ensures that the charge and discharge energy  $s_i$ is within the storage capacity $c_i$. Constraint \eqref{eq:userc2} prevents the negative demand of the user during the peak period.
 
We solve Problem \textbf{UCM} in Proposition \ref{prop:uc}.  We let  $p_i^\Delta$ denote the  price difference   between the peak price  and off-peak price, i.e.,  $p_i^\Delta  \triangleq  p_i^p-p_i^o$.

\begin{prop}[Optimal solution to Problem \textbf{UCM}]\label{prop:uc}
The optimal solution to {Problem \textbf{UCM}} is as follows,
\begin{itemize}
	\item If $\theta_i< p_i^\Delta $, $c_i^*=s_i^*=D_i^p$.
	\item If $\theta_i> p_i^\Delta$,  $c_i^*=s_i^*=0$.
	\item If $\theta_i= p_i^\Delta$,  $c_i^*$ can be any value in $[0,D_i^p]$,  and $c_i^*=s_i^*$.
\end{itemize}
\end{prop}

\textbf{We will show  the proofs of all mathematical results  in the online appendix\cite{project3append}.} Proposition \ref{prop:uc} shows that  users'  optimal storage investment decision has  an \textit{all-or-nothing} property. 
When the storage cost $\theta_i$ is smaller than the price difference   $p_i^\Delta$, user $i$ will invest in the storage capacity equal to his peak demand $D_i^p$, which is also  the charge and discharge energy of the storage. However, if the storage cost $\theta_i$ is higher than the price difference $p_i^\Delta$, no storage will be invested.  Note that this optimal solution is derived under a simplified model of fixed demand across days. However, it can already capture the impact of ToU pricing  on users'  storage investment decisions.\footnote{In reality, considering each user's varying demand across days,  the threshold structure in Proposition \ref{prop:uc} will still hold, although the threshold will depend on the demand distribution. We will generalize the results to the varying demand across day in the future journal version.}

\subsection{Electricity Utility}
  
  The   utility bears the energy supply cost of satisfying users'   demand. We consider a quadratic supply cost, which is commonly used for thermal power plants \cite{touren}.  Usually, the hourly energy, e.g., kWh or MWh, is used to represent the power assuming the constant power within  one hour.  The supply cost for power $p_t$ in hour $t$ is given by $g(p_t)=\alpha p_t^2+\beta p_t+ \gamma$, where the coefficients $\alpha>0$, $\beta>0$ and   $\gamma>0$ are based on practical measurements, such as in  \cite{wu2011generationcost}.  Notice that the power consumption here are aggregated from all users.

Based on the quadratic supply cost, we approximate the power of the peak period  and off-peak  period (with multiple hours) by average energy per hour in these periods,  respectively. For example,  for  the  peak period of 12 hours  with total demand 12 MWh, we use an average demand of 1MW per hour.\footnote{Based the realistic load data of 40 users \cite{diverseload}, the supply cost under  average energy approximation has a  small gap of less than 7\% with the actual supply cost  computed based on actual energy consumption per hour. This motivates us to consider the simplified average energy consumption model. }
Then, for the peak period, if the actual total demand is $L^{p}$ in the system, then the power of each hour is approximated by $L^{p}/H^{p}$. The total  peak period's supply cost $g^{p}$ is then
  \vspace{-1mm}
\begin{align}
g^{p}(L^{p})= \frac{ \alpha }{H^{p}}(L^{p})^2+\beta L^{p}+ \gamma H^{p}.
\end{align}
  Similarly, the  total supply cost for the demand  $L^{o}$ in the off-peak period is 
  \vspace{-2mm}
 \begin{align}
g^{o}(L^{o})= \frac{\alpha}{H^{o}} (L^{o})^2+\beta L^{o}+ \gamma H^{o}.
 \end{align}

\section{Benchmark: Social optimum}
We assume that the regulated utility aims at minimizing the system social cost \cite{contract3}, which   includes users' storage costs and the utility's supply cost.  In this section, we first study  a benchmark where the utility as a social planner  directly decides the optimal  investment and operation of storage in the system, which will be later used in Section \ref{section:contract} to compare with our contracts. In the benchmark,  we will first describe different users types according to different storage costs. Then, we formulate and solve the social-cost minimization problem assuming  the utility knows the aggregate demand of each type.

 \subsection{Type model}  
 We assume  a set of  $\mathcal{K}=\{1,2\ldots,K\}$  storage types with different costs in the market.  The storage cost of type $k$ is denoted by $ \theta^k$.  Multiple users having the same storage cost belong to the same storage type.  We rank the storage types with an increasing order of the storage costs, i.e., $\theta^1<\theta^2<\dots<\theta^K$. Note that the utility does not need to know an individual user's storage cost.

We assume that the utility can estimate the aggregate demand for each storage type, e.g., through surveys among users, historical data of storage incentive program\cite{storageprogram2019utility}, or market share of different storage products \cite{storagemarket2019residential}. We denote the aggregate daily peak demand and off-peak demand for storage type $k$  by $D^{k,p}$ and $D^{k,o}$, respectively.  We consider the following two settings, depending on how much the utility knows  the  information of each type's demand.
\begin{itemize}
	\item  \textit{Complete demand  information}: The utility knows each storage  type's aggregate demands  at the beginning of the investment horizon. 
	\item  \textit{Incomplete demand information}: The utility does not know the exact aggregate demands of each storage type, but  knows the distributions of the demands,  at the beginning of the investment horizon.  
\end{itemize}

Next, we will   consider two benchmarks of  social cost minimization problem under \textit{complete demand information } and \textit{complete demand distribution} of types, respectively.

\subsection{Social cost minimization} 
\subsubsection{Complete demand information} We minimize the social cost in Problem \textbf{SCM} as below, where the utility as a social planner decides the optimal aggregate storage capacity $c^k$  and  charge and discharge  energy $s^k$ for all   users of each type $k$. 

The social cost includes the storage investment cost   $\sum_{k\in \mathcal{K}} \theta^kc^k$ (scaled into one day), the supply cost $g^p$ in the peak period, and the supply cost $g^o$ in the off-peak period. The actual aggregate demand  in the off-peak period is $\sum_{k\in \mathcal{K}} (D^{k,o}+s^k)$ due to the the charged energy $\sum_{k\in \mathcal{K}} s^k$  of all types, and the  actual aggregate demand in the peak period is $\sum_{k\in \mathcal{K}}(D^{k,p}-s^k)$ with  the discharged energy. The constraints follow the same structures as those in users' problem {\textbf{UCM}}, and we just replace the individual users with types.

 \noindent \textbf{Problem SCM: Social Cost Minimization under Complete Demand Information}
 \vspace{-1mm}
 \begin{align}
 \min &\sum_{k\in \mathcal{K}} \theta^k c^k\hspace{-1mm}+ \hspace{-1mm}g^{p}\left(\sum_{k\in \mathcal{K}}(D^{k,p}\hspace{-1mm}-\hspace{-1mm}s^k)\right)\hspace{-1mm}+\hspace{-1mm}g^o\left(\sum_{k\in \mathcal{K}} (D^{k,o}\hspace{-1mm}+\hspace{-1mm}s^k)\right)\notag\\
 \text{s.t.~} &0\leq s^k\leq c^k,\forall k \in \mathcal{K},\\
 &s^k\leq D^{k,p},\forall k \in \mathcal{K},\\
 \text{var:}~ &s^k,c^k,\forall k \in \mathcal{K}.\notag
 \end{align}\par \vspace{-1mm}
 
Problem \textbf{SCM} is a quadratic  programming problem. We will later characterize the structure of its optimal solution.

 \subsubsection{Incomplete demand information}   
 For the incomplete demand information, we model  the peak and off-peak demands  of each type $k$  by  random variables $D^k=(D^{k,p},D^{k,o})$.  
 We focus on the joint distribution of all types' demand. We let $\bm{D}=(D^1,D^2,\ldots, D^K)$ in sample space $\bm{\mathcal{{D}}}$ denote the joint random variable of peak and off-peak demand of all types.  In the benchmark, we  consider the  ideal case that the social planner can decide the optimal storage investment and operation for each realization   $\bm{D}$.  This serves as a lower bound of the best performance  for our  contract. We formulate the expected social-cost minimization problem in Problem \textbf{ESCM}.
 
 \noindent \textbf{Problem ESCM: Expected Social Cost Minimization under Incomplete Demand Information}
 \vspace{-1mm}
 \begin{align}
 Sym^b:=\min ~&\mathbb{E}_{\bm{D}}\Bigg[\sum_{k\in \mathcal{K}}\theta^k {c^k(\bm{D})}\hspace{-1mm}+ \hspace{-1mm} g^{p}\left(\sum_{k\in \mathcal{K}} (D^{k,p}\hspace{-1mm}-\hspace{-1mm}s^k(\bm{D}))\right)\notag\\
&~~~~~~~~~~~~~~~~+g^{o}\left(\sum_{k\in \mathcal{K}} (D^{k,o}+s^k(\bm{D}))\right)\Bigg]\notag 
 \end{align}
  \begin{align}
 \text{s.t.~} 
 & 0\leq s^k(\bm{D})\leq c^k(\bm{D}), \forall k \in \mathcal{K}, \forall \bm{D} \in \bm{\mathcal{{D}}},\\
 & s^k(\bm{D})\leq D^{k,p}, \forall k \in \mathcal{K}, \forall \bm{D}\in \bm{\mathcal{{D}}},\\
 \text{var}:&~ c^k(\bm{D}),s^k(\bm{D}),\forall k \in \mathcal{K}, \forall \bm{D}\in \bm{\mathcal{{D}}} \notag.
 \end{align}
 
Clearly,  Problem \textbf{ESCM} can be decoupled into subproblems for each realization $\bm{D}$, which is equivalent to Problem  \textbf{SCM}. The expected minimum costs of all these realizations will lead to the optimal objective value of Problem \textbf{ESCM}. Thus, we will just focus on  Problem  \textbf{SCM} and analyze it  in next part.

 \subsection{Solution structure of social cost minimization}
 
The optimal solution to  Problem  \textbf{SCM} divides storage types into three classes in terms of  different users' storage investment behaviors. We  show the three-class structure  in Proposition \ref{pro:class}, and present the detailed optimal solution to Problem  \textbf{SCM} in the online appendix\cite{project3append}.
%
%
%

 \begin{prop}[Three classes of storage types] \label{pro:class}
The optimal solution of Problem \textbf{SCM} divides all storage types into three  classes, which are denoted by sets $\mathcal{F}$, $\mathcal{P}$, and $\mathcal{N}$, respectively. 	The storage costs    in Class $\mathcal{F}$ are smaller than the storage costs in Class $\mathcal{P}$, which are in turn  smaller than those in Class $\mathcal{N}$.

 	
 	\begin{itemize}
 		\item Class $\mathcal{F}$: Users of each type $k\in \mathcal{F}$  fully invests in an aggregate storage capacity equal to the aggregate  peak demand, i.e., $c^{k*}=D^{k,p},\forall k\in \mathcal{F}$.
 		\item Class $\mathcal{P}$: There is at most one storage  type  in set $\mathcal{P}$, i.e., $\mid \mathcal{P}\mid\leq 1$. Users of this type partially invests in the aggregate  storage capacity, i.e., it is strictly positive but strictly smaller than  the  peak demand,  $0<c^{k*}<D^{k,p},\forall k\in \mathcal{P}$.
 		\item Class $\mathcal{N}$: Users of each type $k\in \mathcal{N}$  invests in no storage, i.e., $c^{k*}=0,\forall k\in \mathcal{N}$.
 	\end{itemize}
 	
 	Each of the class can be an empty set, with the constraint that  $\mid \mathcal{F}\mid+\mid \mathcal{P}\mid+\mid \mathcal{N}\mid =\mid\mathcal{K}\mid$.

 \end{prop}

 \begin{figure}[t]
 	\centering
 	\includegraphics[width=2.7in]{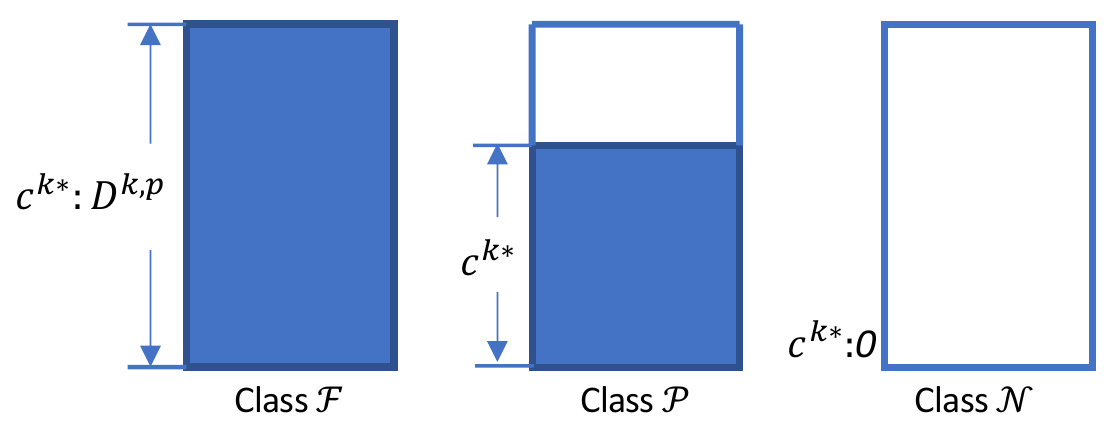}
 	\vspace{-2mm}
 	\caption{\small Illustration of three classes.}
 	\label{fig:ill}
 	\vspace{-4mm}
 \end{figure}

  Proposition \ref{pro:class} classifies storage types into three classes in terms of different optimal storage investment  decisions (that are centrally determined in the benchmark). Note that multiple users may belong to the same type.  If user $i$ belongs to  type $k$ and type $k$ is in any class $\mathcal{F},~\mathcal{P}$, or $\mathcal{N}$, we call that user $i$ is of class $F,P$, or $N$, respectively.  Inspired by the three-class structure, we  next design contracts to guide the  three classes of users to invest in storage  following the social-optimal solution.

\section{Contract Design} \label{section:contract}

We have introduced  the benchmarks of social optimum,  where the utility directly decides the  aggregate storage investment for users  of each type.  In practice,  however, users' storage costs are private information and they will not let the utility  determine the storage investment on their behalf. Hence, we will design contracts  for the utility to incentivize users.  We will design two contracts, where Contract \textbf{TS-C}  achieves the social optimum  for the complete demand information of types, and  Contract \textbf{TS-I}   achieves near social optimum for  the incomplete demand information of types

\vspace{-1mm}

\subsection{Contract design: Complete demand information}
\vspace{-1mm}
 For the complete demand information of types, we will first introduce the contract   and then characterize the conditions to ensure  that the contract can reach social optimum.

 \subsubsection{Contract items} We will design Contract \textbf{TS-C} with three classes (i.e., items), and 
 each item has two parameters:  the peak and off-peak price difference of ToU pricing, and the constraint on each user's storage investment.  The three-item contract is motivated by the social optimum benchmark in   Proposition \ref{pro:class}.  The two parameters are chosen according to the discussions in 
 Proposition \ref{prop:uc} of  users' problem, where users' storage investment decision depends on the comparison between the price difference and the  storage cost.   Next we introduce the contract  in detail.
 

Recall  the utility knows the aggregate demand  of each type under complete demand information. Then, based on the solution the benchmark Problem \textbf{SCM}, the utility knows the optimal aggregate storage investment of each type, and which type belongs to which class. However, the utility does not know  which individual user belongs to which class. Therefore, we design the contract to incentivize  users to reveal the class information. Specifically, for each user of class $x\in \{F,P,N\}$, we design the corresponding contract item $(p_x^{\Delta},\eta_x)$. The parameter $p_x^{\Delta}$ is the  price difference of ToU pricing.\footnote{Note that each contract item specifies only the price difference $p^\Delta$ but not the peak price $p_i^p$ and off-peak  price $p_i^o$. We  assume that the utility can set the peak and off-peak prices for each user $i$ in such a way that, if the user does not invest in storage, his payment is the same across all contract items, i.e.,  $p_{i,F}^p D_i^p+p_{i,F}^o D_{i}^o=p_{i,P}^p D_i^p+p_{i,P}^o D_i^o=p_{i,N}^p D_i^p+p_{i,N}^o D_i^o$, where $p_{i,x}^p$ ($p_{i,x}^o$) denotes the peak (off-peak) price for  user $i$ in class $x$.  As a result, only the price difference affects the user's selection of contract items.} The parameter $\eta_x$ is the maximum ratio of a user's invested storage capacity to his  peak demand, computed as follows  We let  type $b$ be the only type in class $\mathcal{P}$, if $\mathcal{P}$  is non-empty.  
 \begin{itemize}
	\item  $\eta_F=1$: Each user $i$ is required to invest in storage capacity no greater than his  peak demand $D_i^p$.
	\item    $\eta_P=c^{b*}/D^{b,p} \in (0,1)$: Each user $i$ is required to invest in storage capacity no greater than $\eta_P D_i^p $.
	\item  $\eta_N=0$: Each user is not allowed to  invests in  storage.
\end{itemize}

We summarize the design of Contract \textbf{TS-C} as below. Next, we characterize the conditions of the price difference $p_x^\Delta,\forall x\in \{F,P,N\}$  in Theorem \ref{thm3} so as to  ensure the social optimum.
\begin{contract} 
	\renewcommand{\thealgorithm}{}
	\caption{\strut \textbf{TS-C}: Contract for ToU pricing and storage investment under complete  demand information}  
	\label{contract:1}  
	\begin{algorithmic}[1]  \label{contract}
		\STATE \textbf{Information}: The utility knows the information of  each type's aggregate demand.
		\STATE  \textbf{Classification}: The utility solves Problem \textbf{SCM} and divides storage types into three classes $\mathcal{F}$,  $\mathcal{P}$ and $\mathcal{N}$, as stated in Proposition \ref{pro:class}.
		\STATE \textbf{Contract items}: The utility  designs and announces  three-item contract: $(p_x^{\Delta},\eta_x)$, $\forall x\in \{F,P,N\}$:
		\begin{itemize}
			\item $\eta_F=1$, $\eta_P=c^{b*}/D^{b,p}$, $\eta_N=0$.
			\item $p_x^{\Delta}$, $\forall x\in \{F,P,N\}$ satisfying  Theorem \ref{thm3}.
		\end{itemize}
	\end{algorithmic}  
\end{contract}

 \subsubsection{Social-optima conditions}
 We show in  Proposition \ref{prop:op} that  the contract should satisfy two requirements to reach social optimum: (i)  Each user only selects the contract  intended for his class; (ii) Users corresponding to each contract item  invest in storage capacity at the maximum ratio $\eta_x,x\in \{F,P,N\}$.

 \begin{prop}[Social-optimum requirement]\label{prop:op}
 Contract \textbf{TS-C} achieves the social optimum if it induces users' behaviors that satisfy both Requirements (i) and (ii).
 \end{prop}
\noindent   These two requirements will let  the storage investment and operation of all users coincide with the social optimum. Next, we discuss how to achieve such two requirements in detail.

Requirement (i)  is  known as Incentive Compatibility (IC) in Definition 1. We denote the cost of user $i$ after choosing the contract item for class $x$ as $\pi_i(p_x^{\Delta},\eta_x)$.  Later we can show that our contract is always feasible even with such strict inequality.
\begin{defi} [Incentive Compatibility]\label{defi}
	A contract is incentive
	compatible if   user  $i$  of class $x\in \{F,P,N\}$  minimizes his cost by choosing the contract  intended for his class, i.e., for any $y\in \{F,P,N\}$ and $y\neq x$, $\pi_i(p_x^{\Delta},\eta_x) < \pi_i(p_y^{\Delta},\eta_y)$.
 \end{defi}
 \noindent We choose the strict inequality  in Definition \ref{defi} to ensure that a user always chooses the contract item intended for his class  without any ambiguity.  We can achieve Requirement (i)  by satisfying Definition \ref{defi}.

We can achieve Requirement (ii) if we let  the price difference be strictly larger than user $i$'s  storage  cost $\theta_i$ who is expected to invest in storage. This is implied by the all-or-nothing property in  Proposition \ref{prop:uc}.

Then, in Theorem \ref{thm3}, we  characterize the conditions on the  price difference $p^\Delta$  such  that Contract \textbf{TS-C} satisfies  both Requirement (i) and (ii). Here, recall we use type $b$ to denote the only type in class $\mathcal{P}$  (if $\mathcal{P}$ is non-empty). We let Type $a$ be the type with the highest storage cost $\theta^{a}$ in class $\mathcal{F}$, and  Type $c$ be the type with the lowest storage cost $\theta^{c}$ in class $\mathcal{N}$.  If  class $\mathcal{F}$ does not exist, we just set $p_F^{\Delta}=+\infty$ and  $\theta^{a}=0$. If class $\mathcal{P}$ does not exist, we  set $\eta_P=0$ and replace $\theta^{b}$ by  $\theta^{c}$. If the class $\mathcal{N}$ does not exist, we set $\theta^{c}=+\infty$.

 \begin{thm}[Social optima conditions] \label{thm3}
 Contract \textbf{TS-C} achieves the social optimum, if  and only if  the price difference of each contract item  satisfies all the following conditions:
\begin{itemize}
	\item $p_F^{\Delta}$: ${\eta_P \theta^b+(1-\eta_P) \theta^{a} <}p_F^{\Delta} {< (p_P^\Delta-\theta^{b})\eta_P+\theta^{b}.}$
	\item $p_P^{\Delta}$: ${\theta^{b}< } p_P^{\Delta} <\min \left( \theta^{c}, p_F^{\Delta}/\eta_P-\theta^{a}(1-\eta_P)/\eta_P\right).$
	\item $p_N^{\Delta}$: $p_N^{\Delta}\geq0$.
\end{itemize}
Furthermore, there always exist $p_F^{\Delta}$, $p_P^\Delta$,  and $p_N^{\Delta}$ that satisfy the above conditions. 
 \end{thm}
\vspace{-0.5mm}
 
The conditions in Theorem \ref{thm3} ensure that Requirement (i) and (ii) are satisfied. Theorem \ref{thm3} suggests that the price difference  cannot be too higher  or too lower. Intuitively,  a price difference  that is too high (e.g., $p_F^{\Delta}>(p_P^\Delta-\theta^{b})\eta_P+\theta^{b}$) may incentivize users to choose other classes' contract items (e.g., a class-$P$ user may choose the class-$F$ contract item), and  a price difference  that is too low cannot incentivize users to invest in enough storage. 
As long as  the  price difference $p_x^\Delta,\forall x\in \{F,P,N\}$  satisfies the conditions in Theorem \ref{thm3}, the contract will reach the social optimum. Due to the IC requirement, the conditions of  $p_F^{\Delta}$ and  $p_P^\Delta$ are coupled with each other. However, with the strict inequality of storage costs of types, i.e.,    $\theta^{a}<\theta^b<\theta^{c}$,   conditions in Theorem \ref{thm3} are always feasible.

 \vspace{-1mm}

\subsection{Contract design: Incomplete demand information}

In Section \ref{section:contract}.A, we have designed a social-optimal  Contract  \textbf{TS-C} with complete information of each type's aggregate demand. When the utility only knows the distribution of such demands,  we will also propose a three-item contract:  Contract  \textbf{TS-I}, which may not always achieve the social optimum of the benchmark Problem \textbf{ESCM}. We will evaluate the performance of the contract via the performance ratio with the benchmark.

\subsubsection{Contract design}  
In Contract  \textbf{TS-I}, we will still fix the three-class classifications across types. We first decide the optimal partition of the three classes $\mathcal{F}$, $\mathcal{P}$ and $\mathcal{N}$   by minimizing the expected social cost based on the demand distribution of types. Then, we  design the contract following the same way as the case of complete demand information.

Specifically, recall that we let type $b$ be  the only type in  class  $\mathcal{P}$.   For each type $k\in \{1,2,\ldots,b-1\}$, we let the investment ratio $\eta^k=\eta_F=1$. For each type $k\in \{b+1,b+2,\ldots,K\}$, we let the investment ratio $\eta^k=\eta_N=0$.  We denote the investment ratio of type $b$ by $\eta^b$.  
The utility decides the optimal type $b\in\{1,2,\dots,K\}$ in class $\mathcal{P}$ and the optimal ratio  $\eta^b \in [0,1)$ to minimize  the social cost as in the following Problem \textbf{ESCM-C}. The optimal ratio  $\eta^b$ will be later set as $\eta_P$ in the contract. Note that we let $\eta^b \in [0,1]$ instead of  $\eta^b \in (0,1)$ to ensure the feasibility of Problem \textbf{ESCM-C}. If  $\eta^b=0$ or $1$, it will mean that there exists no class $\mathcal{P}$.

\noindent \textbf{Problem ESCM-C:  Expected Social Cost Minimization with Contract under Incomplete Demand Information}
\vspace{-1mm}
\begin{align}
Sym^c:=\min ~&\mathbb{E}_{\bm{D}}\Bigg[\sum_{k=1}^{K}\theta^k \eta^kD^{k,p}\hspace{-1mm}+\hspace{-1mm}g^{p}\left( \sum_{k=1}^{K}(D^{k,p}\hspace{-1mm}-\hspace{-1mm} \eta^kD^{k,p})\right)\notag\\&~~~~~~~~~~~~~~~~+g^{o}\left(\sum_{k=1}^{K} ( D^{k,o}+\eta^kD^{k,p})\right)\notag \Bigg]\\
\text{s.t.~~} 
&\eta^k =1,\forall k \leq b-1,\\
&\eta^b\in [0,1],\\
&\eta^k=0,\forall k \geq b+1,\\
\text{var}: &b \in \mathcal{K}, \eta^b\in [0,1].\notag
\end{align}\par \vspace{-1mm}

Compared with the benchmark Problem \textbf{ESCM},  Problem \textbf{ESCM-C}  fixes the three classes of types independently of  demand's realization.  Thus, there is a gap in the  social cost between Problem \textbf{ESCM-C} and   Problem \textbf{ESCM}. 

Based on the solution to Problem \textbf{ESCM-C},  we design  Contract \textbf{TS-I} following the  way of Contract \textbf{TS-C}. The only difference between Contract \textbf{TS-I}  and Contract \textbf{TS-C} is that the utility solves  Problem \textbf{SCM} in designing Contract \textbf{TS-C}, while solving Problem \textbf{ESCM-C}  in  designing Contract \textbf{TS-I}. We show the detail of Contract \textbf{TS-I} in  the online appendix \cite{project3append}. Furthermore, we show in Theorem \ref{thm4}  that Contract \textbf{TS-I} achievs the optimal solution of  Problem \textbf{ESCM-C}.
\vspace{-0.5mm}
\begin{thm}[Optimality of Contract \textbf{TS-I}]\label{thm4}
	Contract \textbf{TS-I} achieves the optimal solution of Problem  \textbf{ESCM-C}.
\end{thm}
\vspace{-0.5mm}

 Then, it only remains to solve Problem \textbf{ESCM-C}. Problem \textbf{ESCM-C} is a mixed integer quadratic programming. We solve it by an exhaustive search over the  boundary types, which  is efficient since we only compare $K$ results. We show the detailed algorithm in the online appendix\cite{project3append}.


\subsubsection{Performance metric} We evaluate Contract \textbf{TS-I} performance based on the ratio $\kappa$, which is the ratio between the  social cost induced by Contract \textbf{TS-I}  and the minimum social cost of the benchmark Problem \textbf{ESCM}, i.e., 
\vspace{-0.5mm}
\begin{align}
\kappa=\frac{Sym^c}{Sym^b}.
\end{align}\par \vspace{-0.5mm}
Recall that $Sym^b$ is a lower bound of the social optimum under incomplete demand information of types and thus $\kappa\geq 1$. When $\kappa$ is closer to 1, Contract \textbf{TS-I} performs closer to the social optimum. In the next section, we will  use the real data to construct the demand distribution, and we show that the ratio $\kappa$ can be very close to 1 in practice.

\section{Simulation result}

We have shown that for types' complete demand information,  Contract \textbf{TS-C} achieves the  social optimum. For the incomplete demand information, we will further show Contract \textbf{TS-I}'s performance by constructing the type's demand distribution using the realistic data of users' load and solar energy. We show that    Contract \textbf{TS-I}   can lead to the social cost that is very close to the benchmark Problem \textbf{ESCM}.

\subsection{Setup}
\subsubsection{Load  profile}
Based on the Austin Data set\cite{diverseload}, we pick  hourly  load and solar energy generations  of 40 (households) users in one year (with valid data of 361 days).  In Figure \ref{fig:load}(a), we show the aggregate energy profile with seven randomly picked days in one year, where the blue curves and red curves represent the aggregate loads and solar energy generations, respectively. In Figure  \ref{fig:load}(b), we show the aggregate net load (load minus solar energy)\footnote{We let users curtail the surplus renewable energy in simulations.} of seven randomly picked days  in blue curves, with the mean value computed based on the entire year's data in green curve. We construct the users demand distribution based on their net loads profiles of the entire year. 

\subsubsection{Peak  and off-peak periods of ToU pricing} 
\begin{figure}[t]
	\centering
	\hspace{-3ex}
	\subfigure[]{
		\raisebox{-2mm}{\includegraphics[width=1.78in]{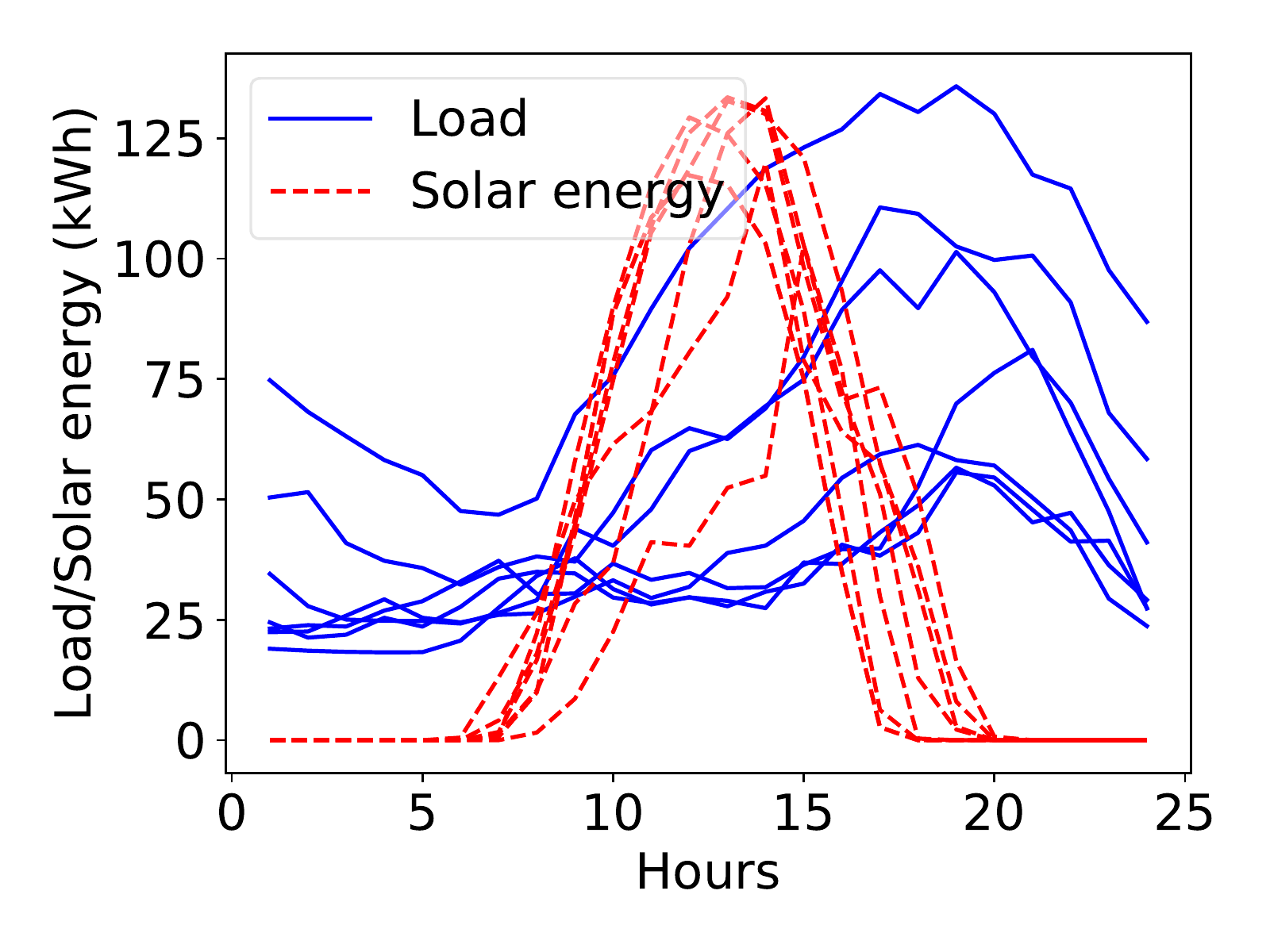}}}
	\hspace{-3ex}
	\subfigure[]{
		\raisebox{-2ex}{\includegraphics[width=1.78in]{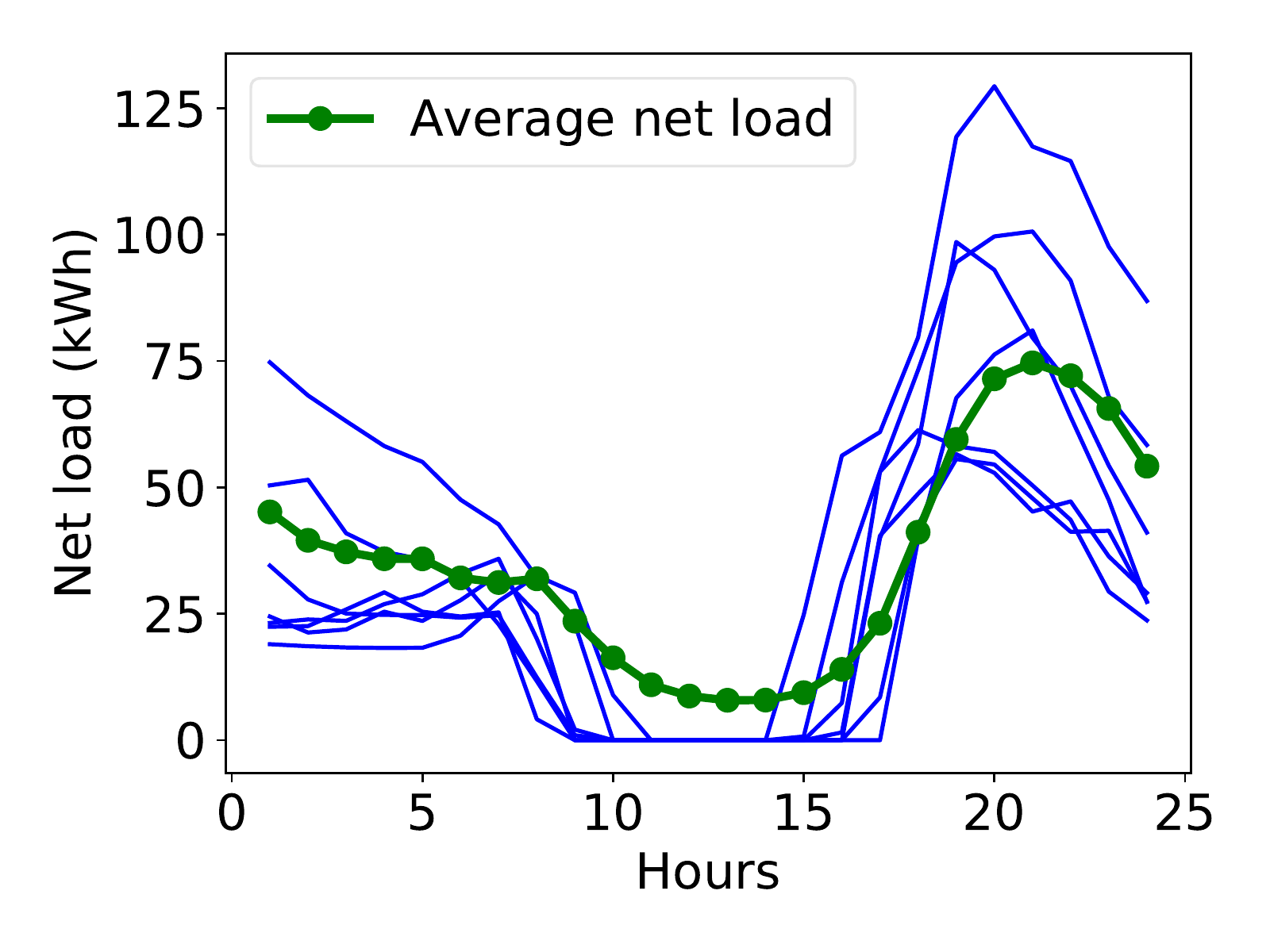}}}
	\vspace{-3mm}
	\caption{\small	(a) Aggregate load /solar energy; (b) Aggregate net load.}
	\label{fig:load}
	\vspace{-4mm}
\end{figure}

Based on the average net load of all users in Figure \ref{fig:load}(b), we empirically set the  peak period  from 18:00 to 00:00 (7 hours), and the off-peak period  from  01:00 to 17:00 (17 hours). 
\subsubsection{Storage cost} We consider  4 storage types with the corresponding (daily) investment costs of  $[\theta_1,\theta_2,\theta_3,\theta_4]$$=\big[\bar{\theta}(1-1.5\lambda^s)$ $,\bar{\theta}(1-0.5\lambda^s),\bar{\theta}(1+0.5\lambda^s),\bar{\theta}(1+1.5\lambda^s)\big]$.  The mean value of  the storage costs is $\bar{\theta}$. The coefficient $\lambda^s$ indicates the diversity of storage costs among types.\footnote {Storage costs can be very diverse. According to  \cite{ralon2017electricity}, the  compressed-air energy storage (CAES) has the cheap capital cost about 53-84\$/kWh with the lifespan of  20-100 years. The Lithium battery's cost can be high. Typically,  Tesla Powerwall's price is 6500\$ for 13.5 kWh with the warranty of 10 years.}  We scale the storage cost  for years evenly into one day, ranging  from 2 \$/MWh (for CAES) to 130 \$/MW (for Tesla Powerwall). In simulations, we  randomly group 40 users into  4 types, and  calculate the mean of  1000 such random groupings of storage types. 

\vspace{-1mm}
\subsection{Performance of the proposed contract}
\vspace{-1mm}
Based on the empirical net-load data, we will show that   Contract \textbf{TS-I} leads to social cost that is very close to the benchmark Problem \textbf{ESCM}, and such a contract significantly reduces  the social  cost compared with no storage investment.

\begin{figure}[t]
	\centering
	\hspace{-3ex}
	\subfigure[]{
		\label{fig:simulation2} 
		\raisebox{-2ex}{\includegraphics[width=1.81in]{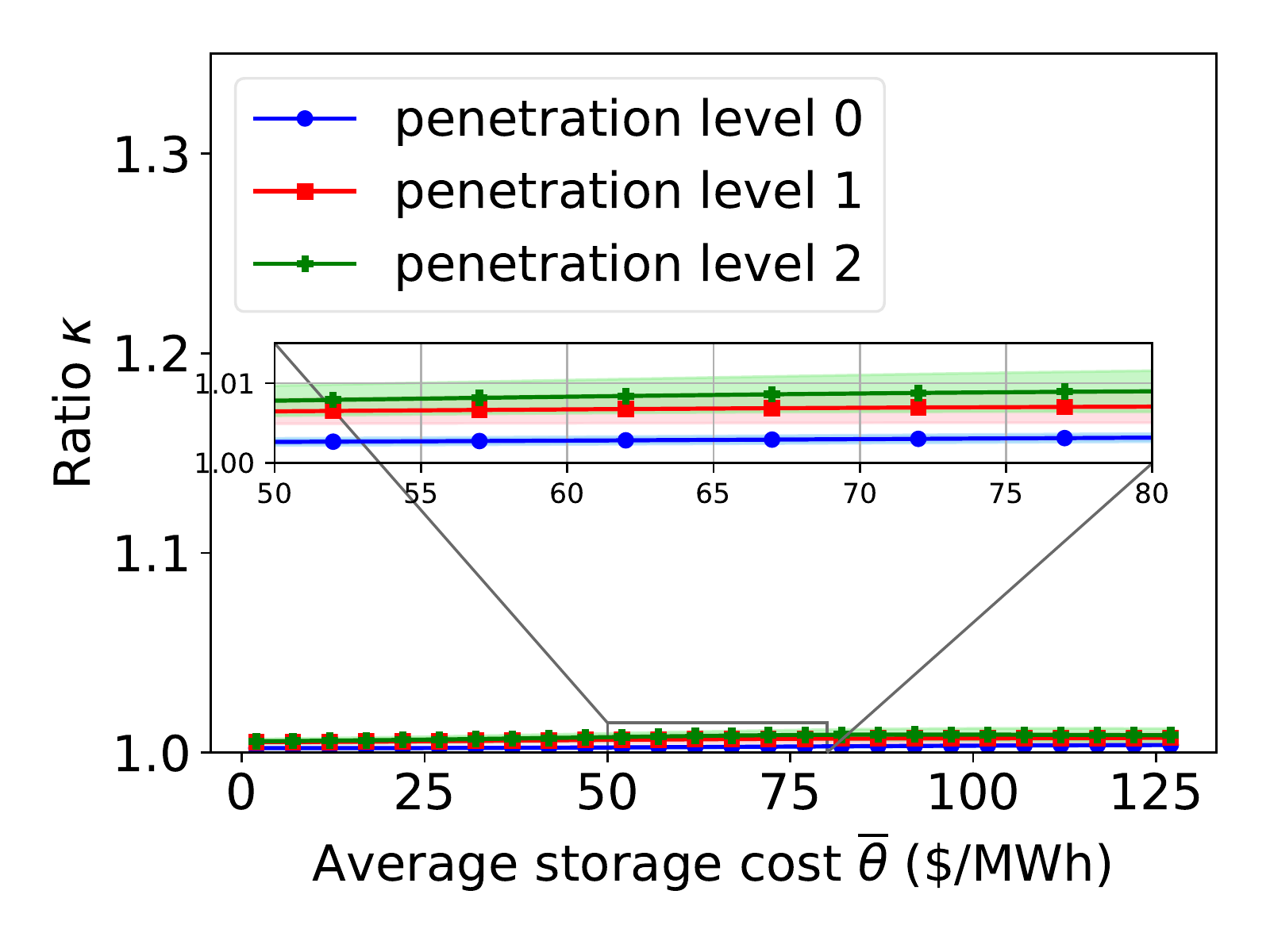}}}
	\hspace{-3ex}
	\subfigure[]{
		\label{fig:simulation3} 
		\raisebox{-2ex}{\includegraphics[width=1.81in]{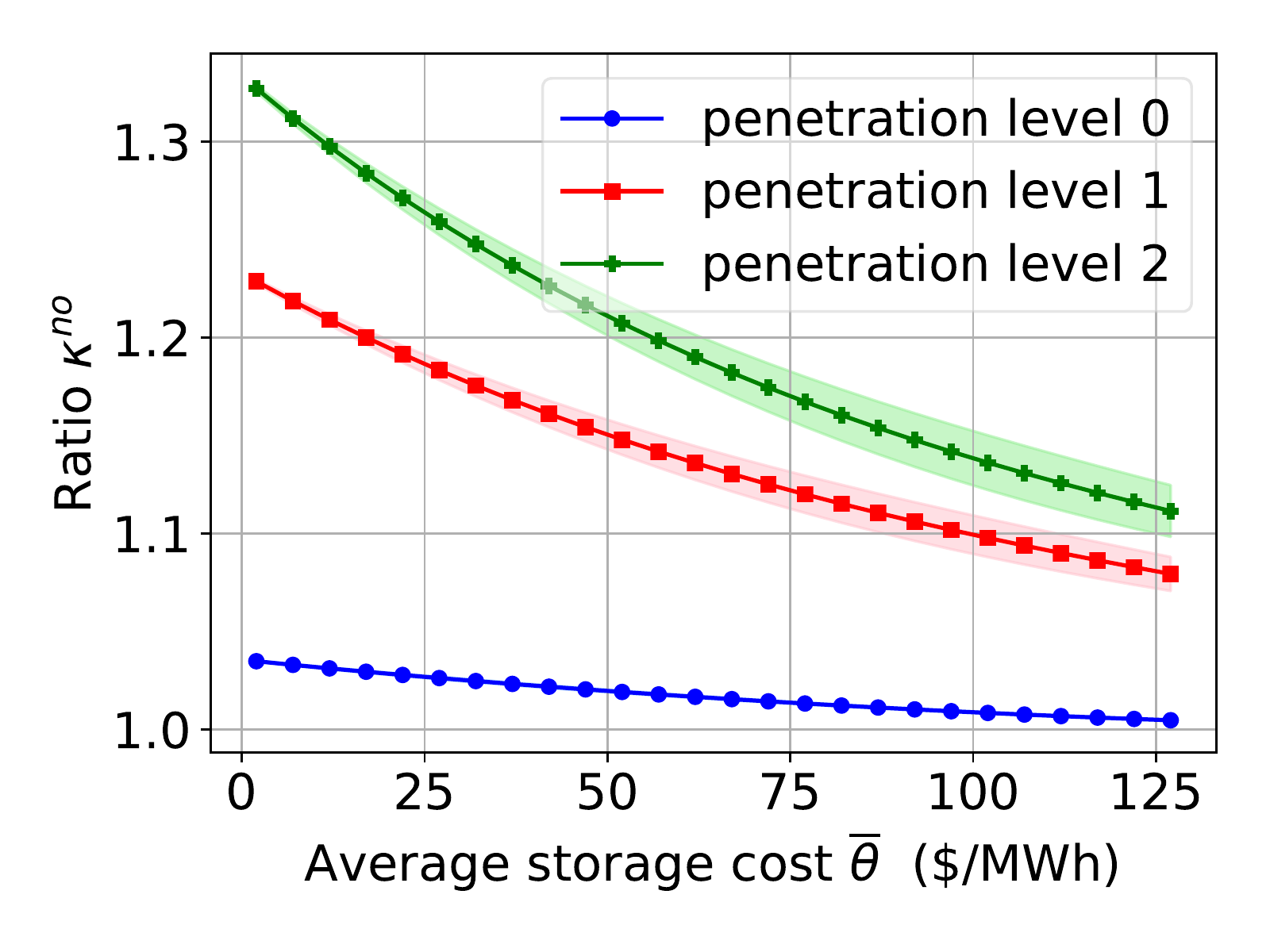}}}
	\vspace{-4mm}
	\caption{\small Contract \textbf{TS-I} performance:  (a) Ratio $\kappa$; (b) Ratio $\kappa^{no}$.  Both are with average storage cost $\bar{\theta}$  at $\lambda^s=1/3$.}
		\label{soc}
	\vspace{-6mm}
\end{figure}

Figure  \ref{soc}(a) shows the average ratio $\kappa$ of all  simulated random type groupings  with different average storage costs $\bar{\theta}$.  Different curves correspond to different penetration levels of solar energy generations, where level 0 means no solar energy, level 1 means the solar energy in the data set, and level 2 doubles the solar energy in the data set (which represents  a future scenario with high renewable penetration).  The shaded regions correspond to the one-standard-deviation ranges  of  all  simulated type groupings.  Based on Figures \ref{soc} (a), we have the following observations. First, Contract \textbf{TS-I}  leads to a social cost that is very close to the benchmark Problem \textbf{ESCM}, and the average  ratio $\kappa$ is always below 1.015. Second,  Contract \textbf{TS-I} is robust among different type  groupings. The standard deviation of the ratio $\kappa$ in the  simulated random type groupings  is  smaller than 0.005.

We define $\kappa^{no}$ as the ratio between the  social costs of  no storage investment and Contract \textbf{TS-I}.   The reason for considering the no storage investment case is that current ToU prices in many places (e.g., most states in US \cite{finkelstein2019residential}) are not high enough to incentivize storage investment.  Figure  \ref{soc}(b) shows the average ratio $\kappa^{no}$  with different  average storage costs $\bar{\theta}$. Different curves correspond to different   penetration levels of solar energy. The shaded regions correspond to the one-standard-deviation ranges  of  all  simulated random type groupings. Compared with the ToU pricing that leads to no storage investment in the system,  Figure   \ref{soc}(b) shows that Contract \textbf{TS-I}  can significantly reduce the social cost, especially when the renewable energy penetration level  is high and the storage cost is low. For example, when the solar energy amount is doubled and the average storage cost is below 15\$/MWh, the social cost can be reduced by over 30\%.

\section{Conclusion}
In this paper,  we   design two contracts of the ToU pricing considering the  storage investment, which deals with the issue of  users' private storage costs.   Both contracts contain only three contract items, which  guides  users of different storage costs to invest in full, partial, or no storage capacity .	 We show that Contract \textbf{TS-C} can   achieve  the  social optimum  under complete information of  the aggregate demand of each type. Via  simulations based on  realistic data, we show that Contract \textbf{TS-I}  can  achieve  near social optimum when the utility only knows the demand distribution of each type.

For the  future work, we plan to consider the more general case where users' demands can vary across days. We expect some of our main conclusions can still hold. For example, users  will still be classified into several classes in terms of the invested storage capacity as a function of their demands, based on which contracts can be designed.

\bibliographystyle{IEEEtran}
\bibliography{storage,bilevel}

\end{document}